{\centerline{\bf{
%%StartAbstract
Discrete Dynamical Models Showing Pattern Formation in
Subaqueous Bedforms.
%%StopAbstract
}}}

\bigskip\bigskip

Prepared for the ``Measures of Complexity and Chaos II,'' 13-15 August 1992,
Bryn Mawr College,
Bryn Mawr, Pennsylvania, USA.
Submitted to International Journal of Bifurcations and Chaos, Special Issue.

\bigskip

{\noindent{
%%StartAbstract
Nicholas B. Tufillaro
%%StopAbstract
\hfil\break}}
Center of Nonlinear Studies, MS-B258\hfil\break
Los Alamos National Laboratory\hfil\break
Los Alamos, NM 87545  USA\hfil\break

\bigskip
%%StartAbstract
Abstract: A new class of ``toy models'' for subaqueous bedform 
formation are proposed and examined. These models all show a similar
mechanism of wavelength selection via bedform unification, and they
may have applications to bedform stratigraphy.
The models are also useful for exploring general issues
of pattern formation and complexity in stochastically driven
far from equilibrium systems. 
%%StopAbstract
\bigskip\bigskip

Running Title: ``Discrete Dynamical Models ...''

\vfill\eject

\bigskip
\baselineskip 20pt

{\leftline{\bf{1. Introduction}}}

The patterns formed (ripples and dunes) on the face of a sandy bed beneath a
flowing
liquid (rivers or tides) are a beautiful example of self-organization in a
geological
system. Faithful models of geological pattern formation (geomorphology) 
can be difficult to specify because
of the nonlinear properties of the material substances (multiphase flows)
as well as the complex and time-dependent boundary conditions present in
the environment. 
Indeed, on the face of it, it seems remarkable that
highly organized rippled bedforms are found to exist with
such great regularity
over a wide range of length scales
and in many diverse environmental
settings [Allen, 1968; Hallet, 1990]. For instance, mud waves on the oceanic
abyssal plane
can be 200 meters high and spaced over 1 kilometer apart. Whereas the size scale
for
sand ripples found beneath tidal currents is of order millimeters 

Here I describe a new class of ``toy models'' which might be
useful in explaining some of the self-organizational mechanisms 
occurring in the formation of subaqueous ripples and dunes. The
cellular automata type models proposed are inspired by
recent developments in physics (the theory
of self-organized criticality [Bak and Chen, 1991; Bak et.\ al.\ 1988]) and
geology
(the formation of eolian ripples [Forrest and Haff, 1992]).
This class of models could also lend itself to
applications involving subaqueous bedform stratigraphy.

In Sec.\ 2 of this paper I briefly review some
features of pattern formation in subaqueous ripples and dunes.
In Sec.\ 3 I provide a (perhaps naive) classification scheme
for the microscopic mechanisms
of sediment transport and then introduce a new class
of cellular automata models to 
mimic these basic sediment transport mechanisms.
In Sec.\ 4 results from
simulations of some of the simpler sediment transport rules
is presented and the common self-organizational principles
exhibited by these models is described. Finally, in 
Sec. 5 I speculate on the relationship between
these microscopic models and continuum (macroscopic)
models for subaqueous bedform evolution.

\bigskip

{\leftline{\bf{2. Basic Subaqueous Bedform Patterns and Transitions.}}}

At least three basic bedforms (patterns) have been identified for
fixed flow and sediment parameters: flat beds, ripples, and dunes.
Several experimental studies [Southard, 1991] have also identified definite
transitions between these basic states as some parameter in the
flow-sediment system is varied, say the flow strength. As the 
flow strength is increased, the simplest observed scenario is a
transition from 
$$
{\rm{flat\ bed\ (low\ flow\ velocity)}} \rightarrow
{\rm{ripples}} \rightarrow {\rm{dunes}} 
\rightarrow {\rm{flat\ bed\ (high\ flow\ velocity)}}.
$$

Ripples can occur at small flow strengths where the
flow over the bedface is laminar. Ripples often exist
as long straight ridges oriented
perpendicular to the flow, although more complex patterns are also common.
The ``ripple index'' (that is, the average ripple wavelength
over the average ripple height) as a rule of thumb tends 
to be of order 10. Ripple geometry is marked by gradual slopes on the
upstream  side of the flow and more steep slopes close to the angle
of repose on the downstream side. Ripples (unlike dunes) are strongly
non-hydrostatic, that is, the ripple geometry does not depend strongly
on the flow depth. In laminar flow regimes, ripple dynamics appears to be
dominated
by bed load transport.

Dunes occur at higher flow strengths where turbulent eddies exist, particularly
in the 
flow of the wake of a dune.
According, sediment in suspension (suspended load) is an important transport
mechanism.
Dunes are strongly hydrostatic. In flume studies, dune height tends to be
one-fifth the flow depth. The ``dune index,'' as a rule of thumb, is of order 5.
Dunes can have a more symmetric geometry than ripples. Like ripples they
can be oriented in long straight ridges perpendicular to the flow, or exhibit
far more complex patterns. Ripples and dunes can coexist in the
same bedform, with smaller ripples traveling up the back of larger dunes.

The idea of forming an analogy between equilibrium thermodynamics phase
diagrams and pattern dynamics is a recurring theme in the study
of complex systems. In the context of subaqueous bedforms, this same
point of view has been advocated by Southard and coworkers [Southard, 1991] who
identify such fundamental parameters as flow speed, sediment size, and flow
depth and then experimentally construct pattern dynamics ``phase diagrams''
showing the relationships between the different bedforms and the
``pattern'' parameters. Extensive analysis of flume and field 
studies shows that the pattern dynamics scenario described above
is, at best, only a very rough approximation to the plethora of patterns
and transitions that occur in a real flow-sediment system.

\bigskip

{\leftline{\bf{3. Discrete Microscopic Models of Sediment Transport.}}}

I find it convenient to separate the
basic microscopic transport processes
into three basic categories or mechanisms: bed load
transport, avalanche transport, and suspended load transport. The
last case, suspended load transport,
consists of the dual processes of scouring and deposition.
Usually, all these transport mechanisms occur only
at a thin layer on the surface of the bedface, a few
grain diameters thick.   

Bed load transport is the local (relative to
the bed surface), short-range movement of sediment
particles tangential to the bedface. It consists of hoping
or rolling motions of individual sediment grains. Typically there is a
short transport distance, say a few grain diameters. This 
transport mechanism exists in both laminar and turbulent flows,
and is driven by the sheer stress generated by the flow
on individual sediment particles.
Presumably, this is the dominate mode of sediment transport
on the upstream side of bed features with a gradual slope.

Avalanche transport can occur on the down stream side of bed 
features with a steep slope close to
the angle of repose. Avalanche processes are quite
visible in experimental flume studies, and in some
parameter regimes are clearly an important method
of sediment transport. However, to the best of my
knowledge, no previous studies have explicitly incorporated
avalanche processes into the modeling of bedform features
and sediment transport. I also think of the avalanche process
as a local mechanism which transports sediment in sudden
bursts. Much of the sediment transported is, again, tangential
to the bedface. Although in a turbulent flow regime, avalanches
can also act to lift large amounts sediment into suspension.

Suspended load transport occurs when turbulent eddies and
bursts lift sediment grains into suspension (scouring). These
sediment grains (in suspension) are then carried by the flow until
they are redeposited (deposition) on the sediment bed. I
think of suspended load transport as a nonlocal mechanism which 
transports sediment particles perpendicular to the bedface.

To model the microscopic dynamics of the
bedface (fluid-sediment interface) I begin by considering
a one-dimensional lattice $j = 1, 2, \ldots, L$ and
associate a sediment height $h_j$ to each site.
For most of the models considered here I also assume
cyclic boundary conditions. The discrete slope
at each site is defined by
$$
s(j) = h(j) - h(j-1) \eqno(1) .
$$
At a discrete time step $t$, the number of particles at
each site can change according to a sediment transport rule
$$
h(j) = B(\cdot) + A(\cdot) + D(\cdot) - S(\cdot), \eqno(2)
$$
where $B$ is a rule describing the bed load transport,
$A$ the avalanche transport, $S$ the suspended load
transport due to scouring, and $D$ the suspended load
transport due to deposition. In general, of course, 
each of these rules can be fantastically complicated, and
can depend on a wide range of variables such as the
local and nonlocal sediment height ($h$), slope ($s$),
an integrated surface area ($I$), and flow parameters
such as the flow strength ($Re$) or sediment size ($Fr$).
For instance, for a complete microscopic theory I would
need to solve the Navier-Stokes equations with time-dependent
boundary conditions, and then within this flow solution
simultaneously solve for the Newtonian motion of individual
sediment particles subject to stresses, flow forces,
gravitational force, buoyancy forces, and so on.

Our main point here, though, is to consider a drastically
simpler class of toy models for the transport
mechanisms in order to explore general features of
pattern formation which I hope 
may hold in a much larger class of transport models.

To this end, I could, for instance, ignore the complications due to the 
flow field by assuming that the flow field (perhaps in a turbulent regime)
stochastically drives the bedface. In this type of toy model
I might imagine that each site randomly ejects a sediment
particle according to some transition probability $p_u$.
Next, this suspended particle is transported a fixed horizontal
distance with each time step. I could think of this adjustable distance
as the flow parameter $Re$. The particle in the flow then gets
deposited with some downward transition probability $p_d$ back on the
bedface.
Meanwhile, the bedface itself can undergo dynamical surface processes
such as avalanches. I like to think of this particular rule as 
a two-level model for sediment transport, the fluid level is 
pictured as a conveyor belt since it simply carries particles
randomly ejected at site $j$ to a downstream site $j+n$ ($n \in Z^+$) in
a few time steps.

As a specific example, consider a stocastically driven, avalanche dominated,
transport model where at each time step $t$, each site obeys the
Kadanoff type avalanche rule [Kadanoff et.\ al., 1989],
$$
A(\cdot) = 
\cases{ {\rm{if}}\ s(i) < -2,
&\ $h(i) \rightarrow h(i) + 2\ {\rm{and}}\ h(i-1)  \rightarrow h(i-1) - 2,$ \cr
{\rm{if}}\  s(i) > +1,
&\ $h(i) \rightarrow h(i) - 1\ {\rm{and}}\ h(i-1) \rightarrow h(i-1) + 1.$ \cr}
\eqno(3) 
$$
Additionally, I consider an initially unpopulated suspended sediment
flow variable $f(j)$
which follows the translation rule at each time step $t$,
$$
f(j) \rightarrow f(j+Re). \eqno(4)
$$
The flow and height variables are coupled by the stochastic rule
$$
S(\cdot) = 
\cases{ 
{\rm{if}}\ (s(j) > 0\ {\rm{and}}\ rand(j,t) > p_u),&
$h(j) \rightarrow h(j) - 1,$\cr
& $f(j) \rightarrow f(j) +1,$
\cr}
\eqno(5)
$$
where $rand(j,t)$ is a random variable between $[0,1]$ and $p_u$ is a fixed
(or variable) transition probability for particle ejection from the bedface, and
$$
D(\cdot) =
\cases{
{\rm{if}}\ f(j) > 0, & $f(j) \rightarrow f(j) - 1,$\cr
&$h(j) \rightarrow h(j) + 1,$\cr}
\eqno(6)
$$
and $D$ is the rule specifying the deposition process.
In this framework a bedface ``equation of motion'' is of the form
$$
s_j(t+1) - s_{j}(t) = A(h_{j-1}, h_{j}) + rand(h_1, h_2, \ldots, h_L, t, t+1) .
\eqno(7)
$$

>From a continuum limit, this bedface dynamic can be viewed as
a kind of (one-sided) nonlinear diffusion process which is driven
stochastically.  

\bigskip

{\leftline{\bf{4. Simulations Showing Self-Organization.}}}

I have explored a large number of sediment transport rules of the type discussed
in the previous section. By fine tuning these rules it is possible to
mimic the (possibly asymmetric) geometry of 
bedform features, such as the angle of an individual
dune on the upstream and downstream side of the flow, and to inhibit
dune growth and size (for instance, by making the transition probability
dependent on the height and slope variables).  The development of 
specific rules for given features could have important applications
for studies of subaqueous ripple stratigraphy. A similar viewpoint
was recently put forth by Forrest and Haff in the context of 
eolian ripples [Forrest and Haff, 1992].

However, our most interesting observation is that the basic pattern
formation processes 
are not very
sensitive to the specific rules chosen. Indeed, finite amplitude
wavelength selection in all the models studied appears to be of the
``bedform unification'' type suggested by Raudkivi and Witte [Raudkivi and

Witte, 1990]. 

This bedform unification process can be seen in Fig.\ 1 which shows
the development of a rippled bedface from a initially flat bed for
the avalanche dominated transport model described in Sec.\ 3.
Due to the stochastic driving, an initially flat bed quickly 
develops small ripples. These ripples then propagate to the 
right with  individual ripple speeds which decrease 
with the ripple size. Thus smaller ripples collide with
and overtake bigger ripples. This collision process is quite
interesting. If the size difference in the ripples is great,
then the smaller ripple is absorbed by the bigger ripple
in one pass. This absorption mechanism is a diffusive like process. 
The speed of the smaller ripple increases as it moves up the back of the 
bigger ripple,
but its characteristic width also
increases so that the smaller ripple appears to diffusively spread out until
it is indistinguishable from the larger ripple.
For ripples of a smaller size difference, the
smaller ripple emerges from the collision with less mass,
some of the mass of the smaller ripple having been absorbed 
by the larger rippler. The smaller ripples than continues for
several more collisions until it is completely absorbed.

A second process is also observed for ripples of a similar size.
Similar size ripples tend to experience a weak repulsive interaction.
If a ripple is slightly smaller it will slowly try to overtake
the larger ripple. However, this (slow) overtaking process involves a net mass
transfer which can often be from the bigger to the smaller ripple.
Thus the smaller ripple grows, slows down, and fails to overtake
the initially bigger ripple.

These two types of ripple interactions seem to be the key elements
present in all models showing the development of regular finite
amplitude rippled bedforms from an initially 
flat bed.\footnote{$^1$}{Similiar results 
have been observed in other types of models for
eolian ripple development by Peter Haff and subaqeous ripple development
by Brad Werner (private communications).}
Qualitative observations of a similar nature have also been
made in experiments studying ripples formed in flumes [Costello and
Southard, 1981].

\bigskip

{\leftline{\bf{5. Continuum Models Based on the Erosion Equation.}}}

A similar type of pattern selection process should also be observable
in continuum models based on the erosion equation [Raudkivi, 1967].
To recall the form of the erosion equation, consider an equation
for the continuity of sediment
$$
{{\partial \epsilon} \over {\partial t}} + \bigtriangledown \cdot 
( {\bf u} \epsilon) = 0 \eqno(8)
$$
where $\bf u$ is the velocity field 
of the fluid
(say, for
instance, a unidirectional flow in the $x$ direction)
and $\epsilon$ is the sediment concentration
field. 
Let $x$ be the downstream horizontal spatial direction and $z$ the vertical
direction.
In this analysis we will not consider the $y$ direction, perpendicular to the
flow.
Further, let $\eta(x,t)$ be the bed height and $z_g(x, t)$ the top of the 
fluid relative to the bed floor. Within the bed ${\bf u} = 0$. Integrating
the continuity equation in the vertical direction and using the
Leibniz rule\footnote{$^2$}{Leibniz rule: 
If 
$$u(x,t) = \int_{a(x,t)}^{b(x,t)} f(x, s, t) ds$$ 
then
$$
u_x(x,t) = \int_{a(x,t)}^{b(x,t)} {{\partial f} \over {\partial x}} ds +
f(x,b,t) {{\partial b} \over {\partial x}} - f(x,a,t) {{\partial a} \over
{\partial x}}
$$ } 
to handle the boundary terms yields the equation
$$
{{\partial \eta} \over {\partial t}} = -{1 \over \epsilon_\eta}
\left[ {\partial \over {\partial t}} \int_\eta^{z_g} \epsilon dz +
{{\partial } \over {\partial x}} \int_\eta^{z_g} {\bf u} \epsilon dz \right]
\eqno(19)
$$
where $\epsilon_\eta$ is the sediment concentration within the bed.
If we let ${\bf q_s} = \int_\eta^{z_g} {\bf u_s} \epsilon dz$ and ignore
the sediment transfer with the water column then we arrive at
the erosion equation
$$
{{\partial \eta} \over {\partial t}} = - {1 \over {\epsilon_\eta}} {{\partial
q_s} 
\over {\partial x}} \eqno(10) .
$$ 
Of course, applying the erosion equation to modeling the bedface dynamics
requires
that some functional relation be established between $\eta$ and $q_s$.
However, given this relationship, we notice that the
erosion equation is a type of kinetic wave equation [Whitman, 1974] that
applies,
for instance, to supersonic shock waves in which a similar behavior of
wave packet interaction is observed as that described in Sec.\ 4.

To explicitly see how these diffusive properties arise from the
erosion equation, assume that $q_s = q_s(\eta, \eta_x)$ depends
locally on the bed height and slope. Then we arrive at diffusive type  equation
of the form
$$
{{\partial \eta }\over {\partial t}} + u_s {{\partial \eta} \over {\partial x}}
= 
D_s {{\partial^2 \eta} \over {\partial x^2}} \eqno(11)
$$
where $u_s = {\partial q_s} / {\partial \eta}$ is the convective speed
of ripples and 
$D_s = -{\partial q_s} / {\partial \eta_x}$  is the diffusion/amplification
factor.

Thus, the discrete models can be viewed as very rough numerical approximations
to
models based on sediment continuity and it may be possible to understand
their behavior by a proper application of kinetic wave theory [Whitman, 1974].
On the 
other hand, since the underlying grain size is discrete, we
like to view the discrete models as no less fundamental than the continuum based
models.

>From a physical point of view I also find these models intriguing because of

the way they can be used to explore the processes by which microscopic disorder
can
create macroscopic order. 

\bigskip

{\leftline{\bf{Acknowledgements.}}}

I acknowledge helpful discussions and correspondence with
Chris Edwards, 
Peter Haff,
A.\ Raudkivi, 
David Rubin,
Roger Samelson, 
John Selker,
Brad Werner,
and 
Kurt Wiesenfeld.

\bigskip
  
{\noindent \centerline{\bf{References.}}}

\medskip

Allen, J.\ R.\ L.\ [1968] {\it{Current Ripples: Their relation to patterns of
water and 
sediment motion}}  (North Holland Publishing Company, Amsterdam).

Bak, P., Tang, C., and Wiesenfeld, K.\ [1988] ``Self-organized criticality,''
Physical Review Letters {\bf{38}} (1), 364-374.

Bak, P.\ and  Chen, K. [1991] ``Self-organized criticality,''
Scientific American 46-53 (January).

Costello, W.\ R., and Southard, J.\ B.\ [1981] ``Flume experiments
on lower-flow-regime bed forms in course sand,'' 
J.\ Sediment.\ Petrol.\ {\bf{51}}, 846-864.

Forrest, S.\ B.\ and Haff, P.\ K.\ [1992] ``Mechanics of wind ripple

stratigraphy,''
Science {\bf{255}}, 1240-1243.

Hallet, B.\ [1990] ``Spatial self-organization in geomorphology:
from periodic bedforms and patterned ground to scale-invariant topography,''
Earth-Science Reviews {\bf{29}}, 57-75.

Kadanoff, L.\ P., Nagel, S.\ R., and Zhou, S.\ [1989]
``Scaling and universality in avalanches,''
Phys.\ Rev.\ A {\bf 39} (12), 6524-6537.

Raudkivi, A.\ J.\ [1967] {\it{Loose Boundary Hydralics}} 
(Pergamon Press, New York).

Raudkivi, A.\ J.\ and Witte, H.-H.\ [1990]
``Development of bed features,'' 
J.\ Hydraulic Eng.\ {\bf 116} (9), 1063-1079.

Southard, J.\ B.\ [1991] ``Experimental determination of bed-form stability,''
Annu.\ Rev.\ 
Earth Planet Sci.\ {\bf{19}}, 423-55.

Whitman, G.\ B.\ [1974] {\it{Linear and nonlinear waves}} 
(John Wiley and Sons, New York).

\end